# A Combinatorial-Probabilistic Diagnostic Entropy and Information

**Henryk Borowczyk,** *Member, IEEE*
*Air Force Institute of Technology, Warsaw, Poland*
*borowczyk@post.pl*

## Abstract

A new combinatorial-probabilistic diagnostic entropy has been introduced. It describes the pair-wise sum of probabilities of system conditions that have to be distinguished during the diagnosing process. The proposed measure describes the uncertainty of the system conditions, and at the same time complexity of the diagnosis problem. Treating the assumed combinatorial-diagnostic entropy as a primary notion, the information delivered by the symptoms has been defined. The relationships have been derived to facilitate explicit, quantitative assessment of the information of a single symptom as well as that of a symptoms set. It has been proved that the combinatorial-probabilistic information shows the property of additivity. The presented measures are focused on diagnosis problem, but they can be easily applied to other disciplines such as decision theory and classification.

*Index Terms*-- entropy, fault diagnosis, information, multi-valued model, uncertainty

## I. INTRODUCTION

Constructing an optimal set of diagnostic symptoms/tests and optimal sequence of gathering/execution thereof is one of the most important problems in engineering systems diagnosis [1] – [7]. Applied optimization method depends on the form of diagnostic model and optimization criterion [1],[2],[5],[8] – [12].

The diagnostic model describes the relationship between a system condition (the set of faults and healthy condition) and diagnostic symptoms [4],[5],[13] – [16].

Most models use binary (good – no-good) conditions and binary (normal – abnormal) symptoms [3],[8],[17]. Better results can be obtained using the qualitative, approximate, and multi-valued models [2],[4],[15],[18],[19]. In [2], it is proved that the length of diagnosis algorithm in the case of multi-valued system conditions and multi-valued symptoms is not larger than the binary algorithm.

Qualitative and multi-valued models can be applied to determine a diagnosis algorithm [1],[2],[19], approximate inference within expert systems [4],[15],[19].

One of the methods of constructing a diagnosis algorithm consists in applying the information-based analysis [2],[3],[8],[16],[18],[20] – [24], that is, description of the system condition uncertainty and the amount of information delivered by means of individual symptoms and sets thereof.

This aim can be reached with the Shannon-introduced quantities: the entropy, and the amount of information [25]. There are some other kinds of entropies [27] that can be considered – Renyi's entropy [26],[28], structural α-entropy [29], and functions $z_\alpha(t)$ [30]. Characterization of information measures (from the information theory point of view) have been extensively discussed in [31],[32].

The abovementioned entropies were introduced for solving information-theoretic problems (e.g. coding). For other problems, different forms of entropy might be more suitable [26],[28],[33]. Therefore, finding the form of entropy best tailored to meet the diagnostic requirements is justified.

This article deals with a multi-valued diagnostic model that exploits the multi-valued system conditions and multi-valued symptoms, where the set of values taken by conditions and symptoms is finite.

The set of desirable properties of the proposed diagnostic entropy is determined by taking the diagnostics point of view into account. The problem is formulated in a system condition-set partition framework [33].

The organization of the article is as follows. In section II, basic assumptions concerning multi-valued diagnostic model are stated. Section III describes information-theoretic, set-partition framework for the diagnosis algorithm designing. It is a starting point for establishing a set of postulated properties of diagnostic entropy that is described in Section IV.

In Section V, the combinatorial-probabilistic diagnostic entropy is introduced and its postulated properties are proved.

The combinatorial-probabilistic diagnostic information of symptoms and sets thereof is defined in Section VI.

## II. ASSUMPTIONS

Further consideration is conducted with the following assumptions formulated and referring to a multi-valued model of the system under the diagnosing.

1. A finite set of the system conditions is determined:

$$E = \{e_i\},\ i = 1,\dots,n \tag{1}$$

Elements $e_i \in E$ can be treated as random events of the type: "a system under diagnosis is in the i-th condition", whereas the whole set $E$ - as a certain event. Conditions might be one-fault or multi-fault ones.

2. The system can remain in one, and only one, of the condition $e_i \in E$ with the probability $P(e_i)$ , and the following holds:

$$\forall_{i=1,\dots,n} P(e_i) > 0, \quad P(E) = 1 \tag{2}$$

Probabilities $P(e_i)$ estimations can be provided by real-world experiment or by using the well-known engineering procedure of the Failure Modes and Effects Analysis [34].

3. Determined is a finite set of symptoms:

$$D = \{d_r\},\ r = 1,\dots,t \tag{3}$$

and a finite set of the symptoms' values

$$A = \{0, \ldots, \lambda - 1\} \tag{4}$$

Multi-valued operator $R(.)$ mapping $E \times D$ into $A$ is crisp, fuzzy, or rough one [2],[18],[35],[36].

If the system is in the condition $e_i$, a value of the symptom $d_r$ is denoted in the following manner:

$$R(d_r / e_i) = g_{ir},\ g_{ir} \in A \tag{5}$$

4. For all the symptoms, the following holds:

$$\mathop{\forall}_{d_r \in D} \mathop{\forall}_{e_i \in E} \mathop{\exists}_{\alpha_{ir} \in E} P[R(d_r/e_i) = \alpha_{ir}] = 1 \tag{6}$$

It means that diagnosis is noiseless [37] or diagnostic model consists of all real-world instances representing class of condition $e_i \in E$.

5. The multi-valued model of the system has been presented in the form of a diagnostic matrix $G$:

$$G = \left[ g_{ir} \right]_{nxt} \tag{7}$$

where:

$$g_{ir} = R(d_r/e_i),\quad g_{ir} \in A \tag{8}$$

The assumed form of the model and the assumptions 1 – 5 offer good representation of many actual systems under the diagnosing [2],[5] – [7],[17],[38],[39].

## III. AN INFORMATION-THEORETIC, SET-PARTITION FRAMEWORK FOR DESIGNING A DIAGNOSIS ALGORITHM

If the system can remain in any of the conditions $e_i \in E$ with the probability $P(e_i)$, the Shannon entropy $H(E)$ thereof is expressed with the following relationship:

$$H(E) = -\sum_{i=1}^{n} P(e_i) \log_\lambda P(e_i) \tag{9}$$

Value $\lambda = 2$ is usually assumed as the base of a logarithm. The general form (9) is more suitable and convenient, if the multi-valued $(\lambda > 2)$ assessment of values of symptoms is a requirement.

The relationship (9) describes the initial uncertainty of the system condition, that is, prior to the selection of any symptom. Assume that any $d_r \in D$ of the symptoms has been selected as the first one. The set of values thereof takes the form (4). Therefore, subsets $E_j(d_r) \subset E$ can be distinguished.

$$\mathop{\forall}_{j=0,\ldots,\lambda-1} E_j(d_r) = \{e_{i_j} : R(d_r / e_{i_j}) = j;\ i_j = 1, \ldots, n_j,\ j \in A\} \tag{10}$$

To simplify the notation, indices have been assigned to the subsets $E_j(d_r) \subset E$, the indices being consistent with suitable values of the symptom $d_r \in D$.

On the assumption (6) that there is full likelihood of the value of the symptom $d_r \in D$, the following are satisfied:

$$
\begin{aligned}
&\text{a)} \quad \underset{\substack{j,l=0,\dots,\lambda-1 \\ j\neq l}}{\forall} E_j(d_r) \cap E_l(d_r) = \varnothing, \quad \bigcup_{j=0}^{\lambda-1} E_j(d_r) = E \\
&\text{b)} \quad \sum_{j=0}^{\lambda-1} n_j = n, \quad \sum_{j=0}^{\lambda-1} p_j = 1, \quad p_j = \sum_{i_j=1}^{n_j} P(e_{i_j})
\end{aligned}
\tag{11}
$$

What follows from (11) is that the selected symptom $d_r$ induces conditions set partition, which is denoted in the following manner:

$$\{E_j(d_r)\} = \{E_0(d_r), \ldots, E_{\lambda-1}(d_r)\} \tag{12}$$

If the symptom $d_r$ takes the value $j \in A$, then according to (10) and (11), the system can find itself in one of the conditions $e_{i_j} \in E_j(d_r)$ with the probability:

$$Q(e_{i_j}) = \frac{P(e_{i_j})}{p_j} \tag{13}$$

Provided that the actual condition $e^*$ of the system has been identified with accuracy to the subset $E_j(d_r)$, the system entropy equals:

$$H(E / E_j(d_r)) = -\sum_{i_j=1}^{n_j} Q(e_{i_j}) \log_\lambda Q(e_{i_j}) \tag{14}$$

Because the symptom $d_r$ can take any value $j \in A$ with the probability $p_j$, the average entropy after having selected the symptom $d_r$ equals:

$$H(E / d_r) = \sum_{j=0}^{\lambda-1} p_j \, H[E / E_j(d_r)] \tag{15}$$

where: $H(E / d_r)$ – the average entropy after having selected the symptom $d_r$.

The amount of information on the system condition contained in the symptom $d_r$ is described as the difference between the entropies $H(E)$ and $H(E / d_r)$:

$$J(d_r) = H(E) - H(E / d_r) \tag{16}$$

As the conditional entropy $H(E / d_r)$ is not larger than the initial entropy $H(E)$, the information $J(d_r)$ is a non-negative quantity:

$$J(d_r) \geq 0 \tag{17}$$

After having selected the first symptom, the conditional entropy $H(E / d_r)$ is usually larger than zero. Therefore, selecting further symptoms proves indispensable. The amount of information entered with the symptom selected in the *k-th* turn is found from the following general relationship:

$$J(d_{(k)}) = H(E / D_{k-1}) - H(E / D_{k-1}, d_{(k)}) \tag{18}$$

where:

$d_{(k)}$ – symptom selected in the *k-th* turn,

$D_{k-1}$ – a set $k-1$ of symptoms selected prior to symptom $d_{(k)}$.

The information-based method consists in subsequent selections of symptoms that contain maximum information on the system condition:

$$J(d_{(k)} / D_{k-1}) = \max_{d_r \in D \setminus D_{k-1}} J(d_r / D_{k-1}) \tag{19}$$

The process is stopped when the remaining entropy is equal to zero or the amount of information of all remaining symptoms equals zero.

One should emphasize that the terms diagnostic entropy and diagnostic information are not the same. Diagnostic entropy is a feature characterizing uncertainty of the system condition and/or complexity of diagnosis algorithm constructing process. Diagnostic information characterises quality of diagnostic symptoms/tests.

The above-presented information-based method of finding a set of symptoms proves to show at least two features of significance:

- the criterion of selecting the symptoms is defined on the basis of a primary term, that is, the entropy, which renders the possibility of gaining coherent measures of the uncertainty of the condition and the amount of information entered by the symptom;
- it uses the property of information additivity – the total amount of information on the system condition, gained after the set of *K* symptoms has been selected, equals the sum of conditional information of individual symptoms:

$$J(D_K) = \sum_{k=1}^{K} J(d_{(k)} / D_{k-1}) \tag{20}$$

However, it appears that the application of the criterion function of selecting the symptoms in the form of the amount of information does not lead to optimal solutions, even in some specific cases, when the set of available symptoms is unlimited.

On the other hand, the idea of information-based approach to the question of optimizing the set of symptoms seems worthy of notice because of the above-mentioned features. A question arises: Is it possible to apply a similar approach on the grounds of some function different than the Shannon entropy, a function that describes the uncertainty of the system condition; and if so, what properties should it then have?

Additivity is the basic feature of the Shannon entropy, which determines the logarithmic form thereof. For two statistically independent systems this feature can be written in the following form:

$$H(E) = H(E_1) + H(E_2) \tag{21}$$

where: $H(E)$ – entropy of a complex system $O$;

$H(E_i)$ – entropy of the system $O_i$.

The two systems are statistically independent if acquisition of information on the condition of one of them does not affect the probability distribution of conditions of the other system.

In diagnostics, the fact of the $O_1$ and $O_2$ systems remaining independent can be interpreted as the symptoms observed with one of the systems remaining independent of the condition of the other system.

Such being the case, the question of condition identification of a complex system resolves itself into two independent questions of condition identification of component systems. Finding a solution to the issue formulated in this way does not require the measure of the uncertainty of the complex system condition to be defined; therefore, there is no need to satisfy the condition of additivity.

What results is a conclusion of great significance to further considerations: the function that describes the uncertainty of the condition of the system under diagnosis does not need the additivity feature in the sense described with (21). This permits to quit the function that shows the logarithmic form, that is, the Shannon entropy. In this way, the first of the above-formulated questions has been affirmatively answered.
The question of properties reads as follows: What properties should a function that describes the uncertainty of the condition show to make it a basis for the information-like approach for optimizing a set of symptoms? And what form should this function take?
The function searched for will be further on called the combinatorial-probabilistic entropy and denoted with $H_B(E)$.

## IV. POSTULATED PROPERTIES OF THE COMBINATORIAL-PROBABILISTIC DIAGNOSTIC ENTROPY

First and foremost, it should be noted that the combinatorial-probabilistic entropy $H_B(E)$ will find its application when designing the diagnosis algorithm. Therefore, this measure should take into account specific features of that process. In particular, it should facilitate quantitative assessment of the partition of the conditions set induced by the selected symptoms.
A set of postulated properties of $H_B(E)$ can be found with its 'conceptual' convergence with the Shannon entropy.
As it follows from (9), probabilities of the system conditions $P(e_i)$ are arguments of the function of entropy. In some indirect way, this function also depends on the cardinality of the conditions set $n = card(E)$. This dependence becomes more evident if the probabilities of all the conditions are $P(e_i) = 1/n, \quad i = 0, \ldots, n$. Such being the case, the function of entropy takes the following form:

$$H(E) = \log_\lambda n \tag{22}$$

However, what most significantly affects the value of entropy is the probabilities of the conditions. This can be illustrated with the exemplary two systems: the sets of conditions thereof as well as probabilities of the conditions are defined in the following way:

$$E = \{e_1, e_2\} \; P(e_1) = P(e_2) = 0,5$$

$$E' = \{e_1', e_2', e_3', e_4', e_5'\}$$

$$P(e_1') = P(e_2') = 0.05, \; P(e_3') = 0,84,$$

$$P(e_4') = P(e_5') = 0,03$$

Entropies of these systems, at $\lambda = 2$, are $H(E) = 1,0$ and $H(E') = 0,947$, respectively. It means that the entropy of a two-condition system is higher than that of a five-condition system. On the other hand, well known is the fact that one symptom is enough to identify the condition of the first system, whereas identification of the second system condition needs at least three symptoms. Therefore, it seems reasonable to postulate that the demanded combinatorial-probabilistic entropy $H_B(E)$ be at the same time the function of the cardinality of the conditions set $n = card(E)$ and probabilities of conditions $P(e_i)$:

$$H_B(E) = f(n, \underline{P}) \tag{23}$$

where: $\underline{P} = \{P(e_i)\}$, $i = 1, \ldots, n$

What else should be expected is the monotonic increase in the value of function (23) with the growth of $n$:

$$f(n, \underline{P}) \geq f(n', \underline{P}') \Leftrightarrow n \geq n' \tag{24}$$

Two other properties result from conditions of setting $H_B(E)$ to zero.

If it is known a priori that the set of system conditions is one-component only ($n = 1$), the condition of the system is then definitely determined and $H_B(E)$ should take value zero:

$$H_B(E)\big|_{n=1} = 0 \tag{25}$$

Another extreme case takes place when the selected symptoms induce partition of the conditions set in the form of one-component subsets $\{\{e_i\}\}$, $i = 1, \ldots n$. It means that all the pairs of conditions have been distinguished by the selected symptoms; therefore, the uncertainty of the condition equals zero:

$$H_B(E / \{\{e_i\}\}) = 0 \tag{26}$$

The demanded function is imposed on with the condition of linearity against the assembly of variables $\underline{P}$ in the following form:

$$f(n, c\,P(e_1), c\,P(e_2), \ldots, c\,P(e_n)) = c\, f(n, c\,\underline{P}) \tag{27}$$

where: $c$ is the nonzero constant.

Because the property (27) has no equivalent among properties of the Shannon entropy, more detailed consideration of resulting consequences seems indispensable. Assume that given is some partition of the conditions set in the form (12). If an actual condition $e^*$ of the system is identified with accuracy to the subset $E_j$, the uncertainty of the condition can be written in the following manner:

$$H_B(E / E_j) = f\left(n_j, \underline{Q}_j\right) \tag{28}$$

where:

$$\underline{Q}_j = \left\{ \frac{P(e_{1_j})}{p_j}, \frac{P(e_{2_j})}{p_j}, \ldots, \frac{P(e_{n_j})}{p_j} \right\}$$

With (27) taken into account, the relationship in (28) takes the following form:

$$H_B(E / E_j) = \frac{1}{p_j} f(n_j, \underline{P}_j) \tag{29}$$

or

$$p_j\, H_B(E / E_j) = f(n_j, \underline{P}_j) \tag{30}$$

where:

$$\underline{P}_j = \left\{ P(e_{1_j}), P(e_{2_j}), \ldots, P(e_{n_j}) \right\}$$

The average uncertainty of the system condition, at the partition in (12) assumed, results from the definition of the statistical mean value:

$$H_B(E/\{E_j\}) = \sum_{j=1}^{\lambda-1} p_j \, H_B(E/E_j) \tag{31}$$

With the following notation introduced:

$$H_B(E_j) = f(n_j, \underline{P}_j) \tag{32}$$

and (30) taken into account , the relationship in (31) can be finally written in the following form:

$$H_B(E/\{E_j\}) = \sum_{j=1}^{\lambda-1} H_B(E_j) \tag{33}$$

The relationship (32) describes the 'contribution' of the *j-th* subset $E_j$ to the average uncertainty of the system condition $H_B(E/\{E_j\})$. Consequently, $H_B(.)$ at each stage of the optimization process can be found by means of the probabilities a priori $P(e_i)$, with no requirement to find conditional probabilities $Q(e_i)$. In other words, imposing condition (27) on the $H_B(.)$, means some reduction in labor demand while carrying out computations to determine the optimal set of symptoms.

## V. THE COMBINATORIAL-PROBABILISTIC DIAGNOSTIC ENTROPY

It follows from the above-presented considerations that the combinatorial-probabilistic entropy $H_B(E)$ of the system under diagnosis should have the following properties:

$$H_B(E) = f(n, \underline{P}) \tag{34}$$

$$H_B(E) \geq H_B(E') \Leftrightarrow n \geq n', \; n = Card(E), \; n' = Card(E') \tag{35}$$

$$H_B(E)\big|_{n=1} = 0 \tag{36}$$

$$H_B(E/\{E_j\}) = \sum_{j=1}^{m} H_B(E_j), \; Card(\{E_j\}) = m \tag{37}$$

$$H_B(E/\{\{e_i\}\}) = 0, \; i = 1, ..., n \tag{38}$$

The form of the function $f(n, \underline{P})$ can be defined with two methods. The first one consists in the formal deduction of the function on the basis of the set of postulated properties. This approach is followed when there is significant proof that there exists only one function showing the preset properties. The second method, usually much simpler, consists in the arbitrary acceptance of a certain form of the function and in proving that it shows the postulated properties.

As there are no limitations put in this article on the number of functions that satisfy conditions (34) – (38) the latter of the methods is applied.

It is worth emphasising that the existence of some 'logical' relation between the combinatorial-probabilistic entropy $H_B(.)$ and the process of planning the diagnostic algorithm should be a reasonable prerequisite for the selection of the form of $H_B(.)$.
Further consideration is based on the following theorem:

Theorem 1
If given are:
a) a finite set of system conditions

$$E = \{e_i\},\ i = 1,...,n \tag{39}$$

b) probabilities of conditions

$$\underset{i=1,\ldots,n}{\forall} P(e_i) > 0 \tag{40}$$

with

$$P(E) = \sum_{i=1}^{n} P(e_i) = 1 \tag{41}$$

then function

$$H_B(E) = \sum_{i=1}^{n-1} \sum_{j=i+1}^{n} (P(e_i) + P(e_j)) \tag{42}$$

which determines the sum of probabilities of all unordered pairs of system conditions shows the postulated properties (34) – (38).

Proof (draft)
To prove Theorem 1 lemmas 1 ÷ 6 are used.

Lemma 1
If assumptions (39) – (41) have been satisfied, the following relationship proves true:

$$\sum_{i=1}^{n-1} \sum_{j=i+1}^{n} (P(e_i) + P(e_j)) = \sum_{i=1}^{n} P(e_i)(n-1) \tag{43}$$

Proof (see APPENDIX)

On the grounds of the Lemma 1, the relationship in (42) can be presented in the equivalent form:

$$H_B(E) = \sum_{i=1}^{n} P(e_i)(n-1) \tag{44}$$

Lemma 2
Equivalence (35) holds for the function (42).

Proof (see APPENDIX)

Lemma 3
For a one-condition system, function (42) takes value equal to zero.

Proof (see APPENDIX)

Lemma 4

If given are:

a) a proper subset $E_j \subset E$, such that:

$$E_j = \{e_{i_j}\}, \quad i_j = 1,...,n_j \tag{45}$$

b) probabilities a priori of the conditions:

$$\underset{e_{i_j} \in E_j}{\forall} P(e_{i_j}) > 0, \quad \sum_{i_j=1}^{n_j} P(e_{i_j}) = p_j \tag{46}$$

then the following equality holds:

$$\sum_{i_j=1}^{n_j-1} \sum_{k_j=i_j+1}^{n_j} (P(e_{i_j}) + P(e_{k_j})) = p_j(n_j - 1) \tag{47}$$

Proof (see APPENDIX)

On the grounds of the Lemma 4, the following notation can be introduced:

$$H_B(E_j) = \sum_{i_j=1}^{n_j-1} \sum_{k_j=i_j+1}^{n_j} (P(e_{k_j}) + P(e_{k_j})) = p_j(n_j - 1) \tag{48}$$

Lemma 5

If given is some partition of the conditions set $\{E_j\},\ j = 1,...,m$ such that:

$$\underset{j=1,\ldots,m}{\forall} \sum_{i_j=1}^{n_j} P(e_{i_j}) = p_j \tag{49}$$

$$\sum_{j=1}^{m} \sum_{i_j=1}^{n_j} P(e_{i_j}) = 1 \tag{50}$$

$$\sum_{j=1}^{m} n_j = n \tag{51}$$

then the following equality holds:

$$H_B(E / \{E_j\}) = \sum_{j=1}^{m} H_B(E_j) \tag{52}$$

Proof (see APPENDIX)

Lemma 6

If the partition of the conditions set is given in the form of one-component subsets $\{\{e_i\}\},\ i = 1,...,n$ , then the uncertainty of the system condition equals zero.

$$H_B(E / \{\{e_i\}\}) = 0 \tag{53}$$

Proof (see APPENDIX)

What has been shown by means of the lemmas 1 – 6 is that function (42) shows the postulated properties (34) – (38), which completes the process of proving Theorem 1.

It comes from the above-presented consideration that function (42), which defines the sum of probabilities of all unordered pairs of conditions of the system under the diagnosing, can be accepted as the combinatorial-probabilistic entropy.

## VI. THE COMBINATORIAL-PROBABILISTIC DIAGNOSTIC INFORMATION

The initial uncertainty of the system condition (prior to the selection of any symptom) is equal to:

$$H_B(E) = n - 1 \tag{54}$$

If any symptom $d_r \in D$, with its set of values as in (4) has been selected as the first one in the sequence, it induces the partition:

$$\{E_j(d_r)\} = \{E_0(d_r), \ldots, E_{\lambda-1}(d_r)\} \tag{55}$$

With full likelihood of the value of symptom $d_r$ satisfied then:

$$\begin{aligned} &\text{a)} \quad \underset{\substack{j,l=0,\ldots,\lambda-1 \\ j\neq l}}{\forall} E_j(d_r) \cap E_l(d_r) = \varnothing, \quad \bigcup_{j=0}^{\lambda-1} E_j(d_r) = E \\ &\text{b)} \quad \sum_{j=0}^{\lambda-1} n_j = n, \quad \sum_{j=0}^{\lambda-1} p_j = 1, \quad p_j = \sum_{i_j=1}^{n_j} P(e_{i_j}) \end{aligned} \tag{56}$$

In compliance with the Lemma 4, the uncertainty of the system condition, after selection of symptom $d_r$ that induces partition (55), equals :

$$H_B(E/\{E_j\}) = \sum_{j=1}^{m} p_j (n_j - 1)$$

As the partition (55) is explicitly defined by means of the symptom $d_r$ that induces it, the above equation can be written as:

$$H_B(E/d_r) = \sum_{j=0}^{\lambda-1} p_j (n_j - 1) \tag{57}$$

It is easy to observe that the uncertainty of the system condition after the selection of the symptom $d_r$ is not greater than the initial uncertainty, that is:

$$H_B(E) \geq H_B(E/d_r) \tag{58}$$

Equality in (58) occurs in the case described with the following condition

$$H_B(E) = H_B(E/d_r) \Leftrightarrow \underset{j\in A}{\exists} (n_j = n \wedge p_j = 1) \tag{59}$$

It means that the value of the symptom $d_r$ does not depend on the system condition, and such a symptom remains useless as far as the condition identification is concerned.
On the grounds of equations (58) and (59), the notion of the combinatorial-probabilistic information of a symptom can be defined.

Definition 1

The combinatorial-probabilistic information of the symptom $d_r$ is equal to the difference in the uncertainty of the condition before this symptom has been selected and the uncertainty left after the selection.

If the symptom $d_r$ is selected as the first one in the sequence, then, according to the Definition 1, the following can be written:

$$J_B(d_r) = H_B(E) - H_B(E/d_r) \tag{60}$$

where: $J_B(d_r)$ is the combinatorial-probabilistic information of the symptom $d_r \in D$.

After substituting (54) and (57) into (60) and conditions of (56) taken into account, the combinatorial-probabilistic information can be presented in two equivalent forms:

$$\begin{aligned} J_B(d_r) &= \sum_{j=0}^{\lambda-1} n_j\,(1-p_j) \\ J_B(d_r) &= \sum_{j=0}^{\lambda-1} p_j\,(n-n_j) \end{aligned} \tag{61}$$

From (58) and (60) it becomes evident that the information $J_B(d_r)$ can take non-negative values only.

$$J_B(d_r) \geq 0 \tag{62}$$

The equation (60) and earlier considerations on the form of the combinatorial-probabilistic entropy $H_B(E)$ give grounds to formulate the following conclusion:

Corollary 1

The combinatorial-probabilistic information $J_B(d_r)$ is equal to the sum of the probabilities of all unordered pairs of conditions distinguishable because of the symptom $d_r$:

$$J_B(d_r) = \sum_{j=0}^{\lambda-2} \sum_{k=j+1}^{\lambda-1} \sum_{i_j=1}^{n_j} \sum_{i_k=1}^{n_k} [P(e_{i_j}) + P(e_{i_k})] \tag{63}$$

This confirms the coherence of the introduced measures of the system condition uncertainty and the information of symptoms.

Selection of one symptom usually does not permit condition identification with the required accuracy; therefore, there is need for selecting further symptoms. If the $d_s \in D$ symptom, with the set of values of the form (4), has been selected as the second one in the sequence, then in each of the subsets $E_j(d_r)$ of the partition (55) induces the following partition:

$$\underset{j=0,\dots,\lambda-1}{\forall} \{E_{j0}(d_r, d_s), \dots, E_{j\lambda-1}(d_r, d_s)\} \tag{64}$$

where:

$$\begin{aligned} &E_{jl}(d_r, d_s) = \{e_{i_{jl}} : R(d_r / e_{i_{jl}}) = j \wedge R(d_s / e_{i_{jl}}) = l, \\ &i_{jl} = 1, \dots, n_{jl}, \} \end{aligned}$$

With full likelihood of the value of the symptom $d_s$ assumed, the following are satisfied:

$$a)\quad \underset{j=0,\dots,\lambda-1}{\forall}\ \underset{\substack{l,k=0,\dots,\lambda-1\\ l\neq k}}{\forall} E_{jl}(d_r,d_s)\cap E_{jk}(d_r,d_s)=\varnothing$$

$$b)\quad \underset{j=0,\dots,\lambda-1}{\forall}\ \bigcup_{l=0}^{\lambda-1} E_{jl}(d_r,d_s)=E_j(d_r) \tag{65}$$

$$c)\quad \underset{j=0,\dots,\lambda-1}{\forall}\ \sum_{l=0}^{\lambda-1} n_{jl}=n_j,\ \sum_{l=0}^{\lambda-1} p_{jl}=p_j,\ \ p_{jl}=\sum_{i_{jl}=1}^{n_{jl}} P(e_{i_{jl}})$$

With the Lemma 4 as the basis and conditions of (56) and (65) taken into account, the relationship that defines the uncertainty of the system condition after having selected both the symptoms, that is, $d_r, d_s \in D$, can be written in the following manner:

$$H_B(E/d_r,d_s)=\sum_{j=0}^{\lambda-1}\sum_{l=0}^{\lambda-1} p_{jl}(n_{jl}-1) \tag{66}$$

On the other hand, the form of the combinatorial-probabilistic information of the symptom $d_s$ under the condition that the symptom $d_r$ has been selected as the first one in the sequence, results from the general Definition 1

$$J_B(d_s/d_r)=H_B(E/d_r)-H_B(E/d_r,d_s) \tag{67}$$

After substituting (57) and (66) into (67), the following is arrived at:

$$J_B(d_s/d_r)=\sum_{j=0}^{\lambda-1} p_j(n_j-1)-\sum_{j=0}^{\lambda-1}\sum_{l=0}^{\lambda-1} p_{jl}(n_{jl}-1)=$$
$$=\sum_{j=0}^{\lambda-1} p_j n_j-\sum_{l=0}^{\lambda-1} p_{jl}n_{jl}-p_j+\sum_{l=0}^{\lambda-1} p_{jl}$$

The conditional combinatorial-probabilistic information of the symptom $d_s$ can be written in two equivalent forms:

$$\begin{aligned} J_B(d_s/d_r)&=\sum_{j=0}^{\lambda-1}\sum_{l=0}^{\lambda-1} n_{jl}(p_j-p_{jl})\\ J_B(d_s/d_r)&=\sum_{j=0}^{\lambda-1}\sum_{l=0}^{\lambda-1} p_{jl}(n_j-n_{jl})\end{aligned} \tag{68}$$

As there is inequality taking place in a way similar to that of (58)

$$H_B(E/d_r)\geq H_B(E/d_r,d_s) \tag{69}$$

therefore,

$$J_\lambda(d_s/d_r)\geq 0$$

The above considerations can be generalized to the question of defining the combinatorial-probabilistic information of the symptom $d_s \in D$ under the condition that the earlier selected set of *k* symptoms $D_k \subset D$, then

$$D_k=\{d_{(1)},d_{(1)},\dots,d_{(k)}\} \tag{70}$$

Assuming that after all the symptoms of the set $D_k$ have been selected, the partition of the conditions set takes the form

$$\{E_j(D_k)\}=\{E_0(D_k),E_1(D_k),\dots,E_{m_k-1}(D_k)\} \tag{71}$$

where: $m_k$ is the power of the family of subsets, which is the partition of the conditions set *E*. The following relationships also need be satisfied:

$$a)\quad m_k \le n \quad b)\quad m_k \le \lambda^k \tag{72}$$

Equality in (72)a is reached when the symptoms from the set $D_k$ induce the (71) partition in the form of one-member subsets, whereas equality in the (72)b, when, after selection of each of the symptoms, the cardinality of the family of subsets, increases $\lambda$ -times. The uncertainty of the system condition after having selected all the symptoms from the set $D_k$ equals:

$$H_B(E / D_k) = \sum_{j=0}^{m_k - 1} p_j (n_j - 1) \tag{73}$$

If the symptom $d_s \notin D_k$ is selected as the next one successively, then it induces partition in the form of (74) in each of subsets of the partition (71) :

$$\underset{j=0,\ldots m_k-1}{\forall} \{E_{j0}(D_k, d_s), \ldots, E_{j\lambda-1}(D_k, d_s)\} \tag{74}$$

where:

$$E_{jl}(D_k, d_s) = \{e_{i_{jl}} \in E : R(d_s / e_{i_{jl}} = l,\ i_{jl} = 1, \ldots, n_{jl}\}$$

With full likelihood of the value of the symptom $d_s$ assumed, the conditions similar to those of (65) are satisfied, and:

$$\underset{j=0,\ldots,m_k-1}{\forall} \sum_{l=0}^{\lambda-1} n_{jl} = n_j,\ \ \sum_{l=0}^{\lambda-1} p_{jl} = p_j \tag{75}$$

The uncertainty of the system condition after having selected all the symptoms from the set $D_k$ and the symptom $d_s \notin D_k$ is equal to:

$$H_B(E / D_k, d_s) = \sum_{j=0}^{m_k-1} \sum_{l=0}^{\lambda-1} p_{jl}(n_{jl} - 1) \tag{76}$$

Using the general Definition 1, the combinatorial-probabilistic information of the symptom $d_s$, under the condition that earlier the set of symptoms $D_k$ has been selected, can be written in the following form:

$$J_B(d_s / D_k) = H_B(E / D_k) - H_B(E / D_k, d_s) \tag{77}$$

After substituting (73) and (76) into (77) and taking (75) into account, the following is arrived at:

$$\begin{aligned} J_B(d_s / D_k) &= \sum_{j=0}^{m_k-1} \sum_{l=0}^{\lambda-1} n_{jl}(p_j - p_{jl}) \\ J_\lambda(d_B / D_k) &= \sum_{j=0}^{m_k-1} \sum_{l=0}^{\lambda-1} p_{jl}(n_j - n_{jl}) \end{aligned} \tag{78}$$

It can be easily noticed that relationships (78) are the generalization of (68); they become identical when $k = 1$, and $m_k = \lambda$.

Another issue of significance is to define the information of the set of $k$ symptoms $D_k \subset D$. In order to do this, the earlier introduced Definition 1 should be generalized to the following form:

Definition 2

The information of the set of symptoms $J_B(D_k)$ is equal to the difference between the initial uncertainty of the condition $H_B(E)$ and the uncertainty left after having selected all the symptoms from the set $D_k$ - $H_B(E/D_k)$:

$$J_B(D_k) = H_B(E) - H_B(E/D_k) \tag{79}$$

After substituting (54) and (73) into (79), the following is arrived at:

$$J_B(D_k) = \sum_{j=0}^{m_k-1} n_j(1-p_j) \tag{80}$$

What comes out from the comparison between (61) and (80) is that both the relationships take identical form in the case $D_k$ is a one-member set.

Assuming, that after having selected all the symptoms from the set $D_k$, the condition uncertainty is non-zero and the symptom $d_s \notin D_k$ has been selected as the next one in the sequence. The joint partition of the set of conditions induced with the symptoms from the set $D_k$ and $d_s \notin D_k$ has taken the form (74), whereas the condition uncertainty is described with the relationship (76). Using Definition 2, the total information of the set of symptoms $D_k$ and the symptom $d_s \notin D_k$ can be presented in the following form:

$$J_B(D_k, d_s) = H_B(E) - H_B(E/D_k, d_s) \tag{81}$$

After substituting (54) and (76) into (81) and transformation with the (75) taken into account, the following is arrived at:

$$J_B(D_k, d_s) = \sum_{j=0}^{m_k-1}\sum_{l=0}^{\lambda-1} n_{jl}(1-p_{jl}) \tag{82}$$

What results from the above-considered issues will be used to prove the Lemma 7, and Theorem 2.

Lemma 7

The total information of the set of symptoms $D_k \subset D$ and the symptom $d_s \notin D_k$ is equal to the sum of the information of the set $D_k$ and conditional information of the symptom $d_s$

$$J_B(D_k, d_s) = J_B(D_k) + J_B(d_s / D_k) \tag{83}$$

Proof (see APPENDIX)

Theorem 2

The information of the set of symptoms $D_K = \{d_k\},\ k = 1,\dots,K,\ D_K \subset D$ equals the sum of conditional information of individual symptoms.

$$J_B(D_K) = \sum_{k=1}^{K} J_B(d_{(k)} / D_{k-1}) \tag{84}$$

Proof (draft):

The proof is carried out with the induction method.

For $K=1$

$$\text{L(84)}=J_B(D_1)=J_B(d_{(1)})$$

$$\text{R(84)}=J_B(d_{(1)}/D_0)=J_B(d_{(1)})$$

that is

$$\text{L(84)}=\text{R(84)}$$

Assuming that (84) proves true for any $1 \le M \le card(D)$

$$J_B(D_M)=\sum_{k=1}^{M} J_B(d_{(k)}/D_{k-1}) \tag{85}$$

After having added $J_B(d_{(M+1)}/D_M)$ to both sides of (85), the following is arrived at:

$$J_B(D_M)+J_B(d_{(M+1)}/D_M)=\sum_{k=1}^{M} J_B(d_{(k)}/D_{k-1})+ +J_B(d_{(M+1)}/D_M) \tag{86}$$

Using the Lemma 7, equation (86) can be brought to the following form (after the transformation of the right side thereof):

$$J_B(D_{M+1})=\sum_{k=1}^{M+1} J_B(d_{(k)}/D_{k-1})$$

This, according to the principle of induction, makes the proof of the theorem complete.
The proof has shown that the combinatorial-probabilistic diagnostic information has the property of additivity. If the amount of information delivered by set of symptoms equals to the entropy value then system condition is identified with necessary accuracy.

## APPENDIX

Proof of Lemma 1 (draft):
After simple transformations on the left side of (43), and changes in summation indices, the following is arrived at:

$$L(43)=\sum_{i=1}^{n} P(e_i)(n-1) \tag{87}$$

The comparison between (43) and (87) shows that:

$$L(43)=R(43)$$ q.e.d.

Proof of Lemma 2 (draft):
On the basis of (44), the following can be written, respectively:

$$\begin{aligned} a)\quad & H_B(E)=\sum_{i=1}^{n} P(e_i)(n-1) \\ b)\quad & H_B(E')=\sum_{i=1}^{n} P(e'_i)(n'-1) \end{aligned} \tag{88}$$

After subtracting the sides of equations (88)a and (88)b, and after the transformation with assumption (41) taken into account, the following is arrived at:

$$H_B(E)-H_B(E')=n-n' \tag{89}$$

The following equivalence directly results from the relationship (89):

$$H_B(E) > H_B(E') \Leftrightarrow n > n' \qquad \text{q.e.d.}$$

Proof of Lemma 3 (draft):

After substituting $n = 1$ into (44) the following is arrived at:

$$H_B(E)\big|_{n=1} = 0 \qquad \text{q.e.d.}$$

Proof of Lemma 4 (draft):

On the grounds of the assumptions (45) and (46), the left side of (47) after suitable transformations, is arrived at:

$$L(47) = p_j(n_j - 1) \tag{90}$$

The comparison between (47) and (90) shows that:

$$L(47)=R(47) \qquad \text{q.e.d.}$$

Proof of Lemma 5 (draft):

The average uncertainty of the condition, for some assumed partition, can be found from the definition of the expected value:

$$H_B(E/\{E_j\}) = \sum_{j=1}^{m} p_j H_B(E/e^* \in E_j) = \sum_{j=1}^{m} p_j(n_j - 1) \tag{91}$$

On the other hand, on the grounds of the Lemma 4, after the summation as related to all *j*, the following is arrived at:

$$\sum_{j=1}^{m} H_B(E_j) = \sum_{j=1}^{m} p_j(n_j - 1) \tag{92}$$

The comparison between (91) and (92) shows that:

$$H_B(E/\{E_j\}) = \sum_{j=1}^{m} H_B(E_j) \qquad \text{q.e.d.}$$

Proof of Lemma 6 (draft):

With the Lemma 5 employed, the uncertainty of the condition can be written in this case in the following form:

$$H_B(E/\{\{e_i\}\}) = \sum_{j=1}^{n} p_j(n_j - 1) \tag{93}$$

After substituting $n_j = 1,\ j = 1,...,n$ into (93), the following is arrived at:

$$H_B(E/\{\{e_i\}\}) = 0 \qquad \text{q.e.d.}$$

Proof of Lemma 7 (draft):

With (80) and (78) as the basis, the following can be written, respectively:

$$\begin{aligned} a)\quad & J_B(D_k) = \sum_{j=0}^{m_k-1} n_j(1 - p_j) \\ b)\quad & J_B(d_s/D_k) = \sum_{j=0}^{m_k-1}\sum_{l=0}^{\lambda-1} n_{jl}(p_j - p_{jl}) \end{aligned} \tag{94}$$

Having summed up (94)a and (94)b by sides,

$$J_B(D_k)+J_B(d_s/D_k)=\sum_{j=0}^{m_k-1} n_j(1-p_j)+$$

$$+\sum_{j=0}^{m_k-1}\sum_{l=0}^{\lambda-1} n_{jl}(p_j-p_{jl})=\sum_{j=0}^{m_k-1}\sum_{l=0}^{\lambda-1} n_{jl}-\sum_{l=0}^{\lambda-1} n_{jl}p_{jl}$$

is obtained. Finally,

$$J_B(D_k)+J_B(d_s/D_k)=\sum_{j=0}^{m_k-1}\sum_{l=0}^{\lambda-1} n_{jl}(1-p_{jl}) \tag{95}$$

What results from the comparison between (82) and (95) is

$$J_B(D_k,d_s)=J_B(D_k)+J_B(d_s/D_k) \qquad \text{q.e.d.}$$